\documentclass[aps,pra,twocolumn,superscriptaddress]{revtex4-1}
\usepackage{graphicx}
\usepackage{caption}
\DeclareCaptionLabelSeparator{dot}{. }
\makeatletter
\def\justified{
	\let\\\@normalcr
	\@rightskip\z@skip \rightskip\@rightskip
	\leftskip\z@skip
	\parindent 0em\relax
	\setlength{\parfillskip}{0pt plus 1fil}}
\DeclareCaptionJustification{justified}{\justified}
\usepackage{subcaption}
\usepackage{amsmath}
\usepackage{braket}
\usepackage{units}
\usepackage{ragged2e}
\usepackage[colorlinks,urlcolor=blue ,citecolor=blue ,linkcolor=blue ]{hyperref}
\usepackage{lipsum}
\usepackage{nameref}
\usepackage{hyperref}
\usepackage{textcomp} %for \textmu
\usepackage{urwchancal}

\renewcommand{\vr}{\boldsymbol r}

\newcommand{\Psit}{\Psi}%{\Psi_{\rm tot}}
\newcommand{\psit}{\psi}%{\psi_{\rm tot}}
\newcommand{\psiu}{\psi_{\rm u}}
\newcommand{\psio}{\psi_{\rm m}}
\newcommand{\psimb}{\bar\psi_{\rm m}}
\newcommand{\Nt}{\bar n}
\newcommand{\Nun}{n_{\rm u}}
\newcommand{\No}{\bar n_{\rm m}}
\newcommand{\nox}{n_{\rm m}}
\newcommand{\nov}{n_{1}}%{\rm ov}}
\newcommand{\ndrt}{n_{0}}%{\rm dr}} %^{\Sigma}}
\newcommand{\fsf}{f_{\rm sf}} %^{\Sigma}}
\newcommand{\nuph}{\nu_{\theta}} %^{\Sigma}}

\newcommand{\Sqo}{S(k,\omega)}

\begin{document}
	
\title{ %Analytical relations between density modulation, superfluid fraction and high-energy scattering response of a supersolid for a new probing scheme.}
{Probing the supersolid order via high-energy scattering: analytical relations among response, density modulation, and superfluid fraction.}}
%What can be learnt on a supersolid from its high-energy scattering response? Analytical considerations.}

\author{L.\,Chomaz}
\affiliation{Institut f\"ur Experimentalphysik, Universit\"at Innsbruck, Technikerstra{\ss}e 25, 6020 Innsbruck, Austria}	
\date{\today}

\begin{abstract}
High-energy scattering spectroscopy is a widely-established technique for probing the characteristic properties of complex physical systems. Motivated by the recent observation of long-sought supersolid states in dipolar quantum Bose gases, I investigate the general relationships existing between the density contrast, the superfluid fraction, and the response to a high-energy scattering probe {of density-modulated states within a classical-field approach.
I focus on the two extreme regimes of \emph{"shallow"} and \emph{"deep"} supersolids, which are of particular interest in describing the phase transitions of the supersolid to a uniform superfluid and an incoherent crystal state, respectively. %the scaling laws relating these three observables in  
Using relevant Ansätze for the fields of dipolar supersolid states in these regimes, I specify and illustrate the scaling laws relating the three observables. This work was first prompted to develop an intuitive understanding of a concomitant study based on experiments and mean-field numerical simulations. Beyond this specific application, this works provides a simple and general framework to describe density-modulated states, and in particular the intriguing case of supersolids. It describes key properties characterizing the supersolid order and highlights new possibilities for probing such properties based on high-energy scattering response. }

\end{abstract}

\maketitle

\section{Introduction}
\label{sec:intro}
Supersolidity is a counterintuitive state of matter, first conceived more than half a century ago~\cite{Penrose:1956,Gross:1957}. The very possibility of a supersolid phase and its surprising properties have raised great debates and triggered  many studies~\cite{Leggett,Andreev:1969,Chester:1970,Kirzhnits:1971cco, Schneider:1971,Balibar:2010,Kuklov:2011,Boninsegni:2012,Yukalov:2020}. A particularly intriguing feature of a supersolid is that it  could be fully phase-coherent (Bose-condensed) while only partially superfluid, its superfluid fraction, $\fsf$, satisfying $0<\fsf<1$~\cite{Leggett}. The experimental search for supersolid states (SSS) had for long remained inconclusive~\cite{Balibar:2010,Kuklov:2011,Boninsegni:2012,Yukalov:2020}. Complementing many efforts focusing on helium, the past decade has seen quantum gases appearing as a promising platform for this search
~\cite{Henkel2010tdr, Cinti2010sdc,Lu2015sds,Li:2017, Leonard:2017,Macia:2016, Cinti:2017, Wenzel:2017, Baillie:2018, Chomaz:2018,Ancilotto:2019,Yukalov:2020}. 

In 2019, breakthrough experiments reported on the observation of SSS in cigar-shaped dipolar quantum gases~\cite{Tanzi:2019, Boettcher:2019,Chomaz:2019}. In these systems, a spontaneous density modulation arises in the axial direction ($x$), via the intrinsic effect of the momentum-dependent inter-particle interactions~\cite{Santos:2003,ODell2003,Lu2015sds,Wenzel:2017,Baillie:2018,Chomaz:2018,Youssef2019,Ancilotto:2019,Tanzi:2019,Boettcher:2019, Chomaz:2019,Yukalov:2020}. A global phase-coherence was additionally shown to be preserved. Such a coexistence of solid and superfluid orders arises thanks to a subtle competition of interactions, at the mean-field and (many-body) beyond-mean-field levels. Calculations performed within an extended mean-field (eMF) approximation show remarkable agreement with experiments. Here, the SSS is described by a classical field, $\Psit(\vr)$, with amplitude modulation along $x$, and its evolution is dictated by an extended Gross-Pitaevskii equation including first order corrections from quantum fluctuations~\cite{Bogoliubov:1947,Lee1957eae,Gross:1961,Pitaevskii:2016,Waechtler:2016,Bisset:2016,Waechtler:2016b,Chomaz:2016,Igor:2016}. 

Several studies on the peculiar dynamical behaviors arising with supersolidity in dipolar gases soon followed, in experiments and using eMF theory~\cite{Natale:2019,Tanzi2:2019,Guo:2019, Hertkorn:2019,Tanzi:2019b,Roccuzzo:2020,Ilzhofer:2019}. Yet, many questions still remain open. Of prime interest are those dealing with the interplay between the fundamental features of the SSS, which are its (partial) superfluid fraction,  
the contrast of the density modulation that characterizes the underlying crystal order, and the possible occurrence of phase variations and fluctuations, which connect to the dynamics, finite temperature, and global coherence of the state. Investigating these features requires developing complementary experimental approaches, which also await further theoretical understanding.

 Scattering spectroscopy, i.e.\,recording the response of a target system to a scattering probe as a function of its energy and momentum, has long been established as an exquisite probe of the structures and properties of quantum many-body states ~\cite{Palevsky:1957,Henshaw:1958,Yarnell:1959,Breidenbach:1969,Pitaevskii:2016}. {Scattering protocols have been developed and applied on a wide variety of systems ranging from nuclear to condensed matter physics~\cite{Palevsky:1957,Henshaw:1958,Yarnell:1959,Taylor1991:DIS, Kendall:1991:DIS, Friedman1991:DIS, Breidenbach:1969,Griffin:1993, GIACOVAZZO:1992}. In the case of the quantum-gas platform relevant for the recent experimental realizations of supersolids, two-photon Bragg spectroscopy is a highly versatile technique for probing the response at momenta and energies ranging from low to high~\cite{Stenger:1999, StamperKurn1999eop, Steinhauer:2002, Ozeri:2005, Richard:2003, Esslinger:2004, Papp:2008, Vale:2010, Fabbri:2011, Vale:2013,Vale:2019, Lopes:2017}. }
 
 In the weak-perturbation regime, scattering spectroscopy connects to the central concept of the spectrum of elementary excitations~\cite{Landau41,Landau47,Feynman1957,Pitaevskii:2016}, dictating the system's dynamical response and thermodynamical properties.  
{In the low-energy regime, the excitation spectrum is sensitive to collective effects and its changes reflects the underlying many-body phases. In the supersolid case, various experimental and theoretical works undertook to characterize the specificity of its low-lying excited modes~\cite{Andreev, Pomeau:1994,saccani:2012, macri:2013,Tanzi:2019,Boettcher:2019,Guo:2019,Natale:2019,Blakie:2020}. A major signature is the occurrence of two phononic branches, connected to the spontaneous breaking of two continuous symmetries.  The high energy scattering regime, achieved by using a probe whose energy is much higher than the characteristic energy of the interaction system and thus well described by free-particle excitations, constitutes a distinct paradigm in which one probes the system's microscopic properties. Whereas it has proven powerful to probe key features such as short-range correlations or beyond-mean-field effects~\cite{Hohenberg:1966,Sears:1982,Sosnick:1989,Zambelli:2000,Pitaevskii:2016,Hofmann:2017,Richard:2003,Esslinger:2004,Papp:2008, Vale:2010, Fabbri:2011, Vale:2013,Vale:2019, Lopes:2017}, high-energy scattering has not been previously applied to SSSs. The present work and Ref.~\cite{Petter:2020} are a first step toward bridging this gap.}

{In the present work, I use simple models relying on a classical-field description to figure out which information is contained in the high-energy scattering response of a SSS. Initiated in relation to the investigations of Ref.~\cite{Petter:2020} based on experiments and numerical simulation, this work provides us with an intuitive understanding of our intricate findings. Beyond this first application, the present work also provides us with a general framework to describe SSSs, sketches general scaling relationships between its most fundamental properties, characterizing its supersolid order, and shows the relevance of high-energy scattering to probe supersolidity.}

The paper is organized as follows. 
Section \ref{sec:general} first introduces our representation of a SSS and decomposes its field in presence of density and phase modulations. The SSS's density contrast, superfluid fraction, and high-energy scattering response are then described, and their expressions within the present representation detailed.  Sections \ref{sec:shallowssp} and \ref{sec:deepssp} explore the behaviors and relations of these three quantities, in the two limiting regimes of \emph{"shallow"} (i.e.\,weakly modulated) and \emph{"deep"} (i.e.\,strongly modulated) supersolids, respectively.  In Sec.\,\ref{sec:shallowssp}, scaling laws are derived from Taylor expansions in powers of the modulated component and exemplified using a sine-modulation Ansatz. 
In Sec.\,\ref{sec:deepssp}, the supersolid is described as an array of (overlapping) drops and the scaling laws extracted from the drop's profile. Conclusions are drawn in Sec.\,\ref{sec:conclusions}.

\section{General Considerations}
\label{sec:general}

The aim of this section is to introduce a general representation of a SSS (Sec.\,\ref{subsec:decomp}), and to specify within this framework the expressions of the state's main characteristics (Sec.\,\ref{subsec:ssschar}), namely its density contrast [Eq.\,\eqref{eq:Cnop}] and superfluid fraction [Eq.\,\eqref{eq:fsfut}], as well as  of
its response to a high-energy scattering probe [Sec.\,\ref{subsec:highscatt}, Eq.\eqref{eq:Sq}]. These general expressions are the basis to the derivations of  Secs.\,\ref{sec:shallowssp}, \ref{sec:deepssp}. 

\subsection{Supersolid state's decomposition}
\label{subsec:decomp}
In the present model, a SSS is depicted by a classical field,  
$\Psit (\vr)$. This is particularly justified in the context of dipolar supersolids, given their successful description within an eMF theory~\cite{Ancilotto:2019,Tanzi:2019, Boettcher:2019,Chomaz:2019,Natale:2019,Tanzi2:2019,Guo:2019, Hertkorn:2019,Tanzi:2019b,Roccuzzo:2020}. The density modulation is assumed to arise along one direction of space, $x$, as relevant for the cigar-shaped configurations of the above-cited experiments. For simplicity, I consider a system confined in a uniform box whose size along $x$ is denoted $L$ with periodic boundary conditions at the interval $[0,L]$ bounds. I focus on the system's properties along the modulation direction and disregard the system's behavior in the $yz$-plane by assuming the spatial dependencies to be separable. Then, the wave function decomposes as $\Psit (\vr)=\psit(x)\xi(y,z)$ with $\int |\xi(y,z)|^2dy\,dz =1$~\cite{Blakie:2020,Pitaevskii:2016,Salasnich:2002}, and integrating out the $yz$ dependence of $\Psi(\vr)$ yields an effective one-dimensional model on $\psit(x)$. 

The SSS's crystal order implies that the amplitude $|\psit(x)|$ is modulated. 
A SSS is standardly described considering a fully-coherent many-body ground state, which corresponds 
to a single classical field with amplitude modulation and uniform phase~\cite{Gross:1957, Leggett, Pomeau:1994,Josserand:2007, Sepulveda:2008, Kirzhnits:1971cco, Ancilotto:2019, Boettcher:2019,Chomaz:2019,Natale:2019,Tanzi2:2019,Guo:2019, Hertkorn:2019,Tanzi:2019b}. We here extend the state's description beyond this case in order to encompass several physical effects that are relevant for dipolar SSSs. We here allow for spatial modulations of $\psit(x)$ not only in amplitude but also in phase, which enable to capture the possible existence of excitations in the state~\cite{Pitaevskii:2016}. 

Excitations can occur in SSSs for two main reasons, namely dynamics and statistical fluctuations.
Dynamics is relevant for dipolar SSSs since, in order to reach such states in experiments, dynamical tuning of parameters and crossing of phase transitions are performed~\cite{Wenzel:2017,Boettcher:2019,Tanzi:2019,Chomaz:2019,Tanzi2:2019,Guo:2019,Ilzhofer:2019}.  
The resulting coherent dynamics yields deterministic phase patterns, i.\,e.\,which are reproducible from one realization of  $\psit$ to the other. Furthermore, statistical fluctuations are important   due to both the enhanced role of quantum fluctuations in the dipolar SSS~\cite{Igor:2016,Chomaz:2016,Waechtler:2016,Waechtler:2016b,Bisset:2016,Wenzel:2017,Tanzi:2019, Boettcher:2019,Chomaz:2019} and the finite temperature of experimental systems. The present classical-field model can be used to approximately account for statistical-fluctuation effects~\cite{Blakie:2008}. This is achieved by performing ensemble averages over sets of $\psit$ including phase and amplitude modulations of random character, and appropriately sampling the phase space.
For instance, one can add thermal and quantum noise in the weakly correlated regime by superimposing excited modes on this initial state, with the complex amplitudes of these modes drawn from the appropriate (Wigner) distribution~\cite{Blakie:2008}.
I note that the effect of such fluctuations will be particularly relevant in the "deep" supersolid regime of Sec.\,\ref{sec:deepssp}, dictating the transition to an incoherent crystal state. In the following, I will give the expressions of the state's properties for a single modulated field. A simple extension to fluctuating configurations is provided via the above-discussed statistical sampling.

To analyze the SSS's properties, I use a convenient decomposition of  the axial field, $\psit(x)$, in two components: (i) a spatially uniform (u) component, $\psiu$, corresponding to the spatial mean of the field, and (ii) a modulated (m) component, $\psio(x)$, accounting for spatial modulations of the field's phase and amplitude,
\begin{eqnarray}
\label{eq:decpsi}
\psit(x) &\equiv& \psiu+\psio(x),\\
\label{eq:psiu}
\psiu &\equiv& \int_0^L \psit(x)\frac{dx}{L}.
\end{eqnarray}
It follows from this definition that $\psio$ spatially averages out, 
\begin{eqnarray} 
\label{eq:psio}
\int_0^L \psio(x)dx =0.
\end{eqnarray}
The global phase of $\psit(x)$ is chosen such that $\psiu$ is real~\footnote{Starting from an arbitrary phase choice, the appropriate correction would be made by subtracting a uniform phase matching the argument of $\int \psit(x)dx$.}. 
The above decomposition is particularly convenient for our purposes, since, as we will see in Sec.\,\ref{subsec:highscatt} [Eq.\eqref{eq:Sq}], the amplitude of the high-energy scattering response on resonance, which is our observable of interest, is uniquely determined by $\psiu$, see also Apendix \ref{appendix:NuNt}. 

In the remainder of this section, I detail how the SSS's properties read within the decomposition of Eqs.\eqref{eq:decpsi}-\eqref{eq:psio}. 
In this aim, a first step is to write the spatial density distribution. This is 
\begin{eqnarray}
\label{eq:nx}
n(x)=|\psit(x)|^2 =\Nun+|\psio(x)|^2+2\psiu {\rm Re}[\psio(x)],
\end{eqnarray}
with $\Nun$ being the density of the spatially uniform component, $\Nun=\psiu^2 $. 
The cross term $\psiu \psio(x)$ appearing in Eq.\,\eqref{eq:nx} implies that the spatial density cannot be generally decomposed into independent contributions from the uniform and modulated fields.
In contrast, Eq.\,\eqref{eq:psio} yields the simple decomposition of the spatial mean density, 
\begin{eqnarray}
\label{eq:nt}
\Nt \equiv \int  n(x)\,\frac{dx}{L} = \Nun + \No,
\end{eqnarray}
where $\No$ is the mean density in the modulated field,
\begin{eqnarray}
\label{eq:no}
\No =\int |\psio(x)|^2\,\frac{dx}{L}. 
\end{eqnarray}
We also introduce the effective profile 
\begin{eqnarray}
\label{eq:nop}
\nox(x)=n(x)-\Nun =|\psio(x)|^2+2\psiu {\rm Re}[\psio(x)],
\end{eqnarray}
which can take negative values. Despite that $\nox(x)$ differs from $|\psio(x)|^2$, these two profiles have the same average value, $\No$.

\subsection{Supersolid's characteristics: density contrast and superfluid fraction}
\label{subsec:ssschar}

Supersolids are states of matter where both crystal and superfluid orders coexist. To quantify these two distinct characters, two physical observables are needed. Here, I specify such observables and express them within the model~\eqref{eq:decpsi}.

The crystal order of a SSS relates to the amplitude modulations in its field, $\psit(x)$. Their strength is characterized via the contrast of the modulations induced in the density profile~\cite{Boettcher:2019,Tanzi:2019,Chomaz:2019,Ilzhofer:2019, Blakie:2020}, $C$, defined as
\begin{eqnarray}
\label{eq:C}
C=\frac{\max(n)-\min(n)}{\max(n)+\min(n)},
\end{eqnarray}
where $\max$ ($\min$) denotes the maximum (minimum) over space (i.e.\,$x\in [0,L]$). By definition, $C=0$ ($C=1$) is equivalent to $\psit(x)$ having a uniform amplitude (vanishing somewhere). 

The superfluid order is standardly quantified via the superfluid fraction, $\fsf$~\cite{Landau41,Pitaevskii:2016}. In a seminal work~\cite{Leggett}, Leggett explicitly related the superfluid character of a fully coherent state to the spatial dependence of its wave function. Leggett's derivation is based on the concept of non-classical rotational inertia: when a system is put under rotation, its energy changes; the ratio between the quantum and the classical expectations for this energy change is given by the superfluid fraction. 
Thus, by evaluating the kinetic energy of a fully coherent SSS under the assumption of a stationary flow and twisted boundary conditions, Leggett found an upper bound for the supersolid's superfluid fraction. This upper bound was further shown to be saturated in the case of a one-dimensional (axial) density modulation~\cite{Sepulveda:2008}, as considered in the present work and relevant for the dipolar-SSS experiments. Therefore, I define $\fsf$ using Leggett's formula, \begin{eqnarray}
\label{eq:fsf}
\fsf \equiv  L^2\left(\int\displaylimits_0^L dx|\psit(x)|^2\int\displaylimits_0^L \frac{dx}{|\psit(x)|^2}\right)^{-1}.
\end{eqnarray}
Equation \eqref{eq:fsf} evaluates the state's superfluid fraction for $\psit(x)$ real (i.e.\,for a steady coherent state)~\cite{Leggett, Sepulveda:2008,Josserand:2007}.  When $\psit(x)$ presents phase modulations, as can be considered in this work (see Sec.\,\ref{subsec:decomp}), we note that Eq.\,\eqref{eq:fsf} 
is still mathematically defined, yet without direct physical meaning.  Equation \eqref{eq:fsf} implies that, if $\psit(x)$ has a uniform amplitude (cancels somewhere), then $\fsf=1$ ($\fsf=0$). When uniform, a fully coherent state is entirely superfluid, whereas it loses its superfluid character if its density vanishes somewhere in space.

Using the decomposition of the density profile of Secs.\,\ref{subsec:decomp} [Eqs.\eqref{eq:nx}-\eqref{eq:nop}], $C$ and $\fsf$ rewrite
\begin{eqnarray}
\label{eq:Cnop}
C&=&\frac{\max(\nox)-\min(\nox)}{2\Nun + \max(\nox)+\min(\nox)},\\
\label{eq:fsfut}
\fsf &=&  \left(\int\displaylimits_0^L \frac{\Nt }{\Nun+\nox(x)}\,\frac{dx}{L}\right)^{-1}.
\end{eqnarray}
Without making additional assumptions, no general formula relates $C$ and $\fsf$ to the averages $\Nun$ and $\No$. In the following sections, such relations will be extracted in limiting cases or by relying on wave-function Ansätze. 

\subsection{High-energy scattering response}
\label{subsec:highscatt}

{As introduced in Sec.\,\ref{sec:intro}, the focus of the present work is to understand which properties of many-body quantum supersolid states are probed via scattering spectroscopy, especially focusing on the high-energy regime. Considering a scattering probe of momentum $\hbar\boldsymbol{k}$ and energy $\hbar \omega$, the regime of high-energy scattering is defined, within a classical-field picture, by comparing $\hbar\omega$ and $\hbar k^2/2m$ ($m$ is the particle's mass) to the characteristic energy scales of the interactions of particles within the state $\psit(x)$ and scattered to a state of momentum $\hbar \boldsymbol{k}$~\cite{Hohenberg:1966,Zambelli:2000,Pitaevskii:2016,Hofmann:2017}, see also Appendix \ref{appendix:Dirac}. In the regime where the probe energies are large compared to the interactions, the system's excitations  are well approximated by free-particle states, mostly unaffected by interaction effects, justifying an "impulse approximation". For ultracold quantum gases, this regime is readily achievable based on two photon Bragg scattering~\cite{Richard:2003,Esslinger:2004,Papp:2008, Vale:2010, Fabbri:2011, Vale:2013,Vale:2019, Lopes:2017,Petter:2020}.}

This section expresses the resonant response of a SSS to such a  high-energy high-momentum scattering probe, and relates it to the decomposition (\ref{eq:decpsi}-\ref{eq:psiu}). Exploiting the geometry described in Sec.\,\ref{subsec:decomp}, the discussion is restricted to axial excitations with momenta along $x$ and $|\boldsymbol{k}|=k_x \equiv k$. I additionally disregard the effect of transverse excitations and assume that the excited modes' wave functions decompose into $\varphi_j(x)\xi(y,z)$ for all modes $j$ responding to the probe.   
In the high-energy regime, the excited state $j$ of momentum $\hbar k_j=h j/L$ has a wave function $\varphi_j(x)\approx e^{ik_j x}/\sqrt{L}$ well approximated by a plane wave, and an energy $\hbar\omega_j$ dominated by the kinetic term $\hbar^2 k_j^2/2m$.

{In the linear regime, the response of a physical system to a scattering probe of momentum $\hbar\boldsymbol{k}$ and energy $\hbar \omega$ is given by the dynamic structure factor (DSF),  $S(\boldsymbol{k},\omega)$~\cite{Pitaevskii:2016,Pines:1966,vanHove1954}. It connects to the expectation of the squared (equal-time) density fluctuations at momentum $\hbar k$. Focusing on the high-$k$ regime where interaction effects are negligible,  this is simply given by $\sum_j\big|\int dx\,\textrm{e}^{ikx}\varphi_j^*(x)\psit(x)\big|^2$~\cite{Hohenberg:1966,Zambelli:2000,Pitaevskii:2016,Hofmann:2017}. The DSF in this so-called impulse approximation then reads
\begin{equation}
    \Sqo=\sum_j \frac{\tilde n(k-k_j)}{L}\delta(\hbar\omega-\hbar\omega_j),
    \label{eq:SqIA}
\end{equation}
where $\delta(\cdot)$ is the Dirac delta function, and $\tilde n(k)$ is the momentum distribution corresponding to $\psit(x)$, i.e.\,$\tilde n(k) =|\tilde \psit(k)|^2$ with %$\tilde \psit(k)$ being the Fourier transform of the $\psit(x)$,  
$\tilde \psit(k) = \int dx\,\textrm{e}^{ikx}\psit(x)$.}

Probing the system on resonance with a state $j$, i.e.\,at $k=k_j$ and $\omega=\omega_j$, the response is given by
\begin{eqnarray}
\label{eq:SqIAr}
  S(k_j,\omega_j)  =  \frac{m \tilde n(0)}{2\pi \hbar^2k_j} \equiv S  \frac{m L}{2\pi \hbar^2 k_j},
\end{eqnarray}
The factor $\frac{mL}{2\pi \hbar^2 k_j}$ appearing in Eq.\,\eqref{eq:SqIAr} is the density of excited states at the energy $\hbar \omega_j$. In general, the density of excited states could depend not only on the excitation parameters but also on the state's properties.
Yet, at large $k_j$, this factor is purely given by the excitation momentum and does not depend on $\psit$, see Appendix \ref{appendix:Dirac}. In the following, I focus on the dimensionless factor $S$ that accounts for all information on the state's wave function $\psit$ contained in the high-momentum DSF. This factor being simply $S=\tilde n(0)/L$   implies that the resonant scattering response directly probes the system's momentum distribution at $k=0$. The component $\tilde n(0)$ gives the spatial integral of the one-body density matrix, $G_1(x,x')=\langle \hat\psi^\dagger(x)\hat\psi(x')\rangle$,  $\tilde n(0)= \int dx dx' G_1(x,x')$. Here $\hat\psi(x)$ is the field operator, annihilating a particle at position $x$, $\langle \cdot \rangle$ is the state average, and $n(x)=G_1(x,x)$. The matrix $G_1(x,x')$ characterizes the correlations in the system and its long-distance behavior informs on the system's coherence (i.e.\,off-diagonal long-range order).  We note that, because the SSS of interest here is intrinsically density modulated, thus nonuniform, $\tilde n(0)$ provides information not only on the coherence properties of the state but also on the modulations of the order parameter. 
 
Using the general decomposition of Sec.\,\ref{subsec:decomp}, which yields by definition $\tilde \psit(0)=L\psiu$ [Eq.\,\eqref{eq:psiu}], the resonant high-energy DSF of a modulated state is simply given by   
\begin{eqnarray}
\label{eq:Sq}
S = \Nun L= (\Nt-\No)L,
\end{eqnarray}
i.e.\,by the density of the uniform component. It is interesting to note that this response is reduced compared to that of a uniform state of identical  mean density $\Nt$. The reduction factor, given by the fraction of atoms encompassed in the modulated field [Eq.\,\eqref{eq:no}], arises from  both density and phase modulations, see also Appendix \ref{appendix:NuNt}. 

{Finally, we note that the DSF on resonance is a convenient observable that one can readily characterize from scattering spectroscopy measurements. In such a measurement, after applying a scattering probe of momentum $\hbar k$ and energy $\hbar \omega$ for a time $\tau$, one typically extracts the fraction of particle excited to the momentum $\hbar k$ or equivalently the total momentum transferred to the system. In cold-atom experiments, this is achieved by imaging the cloud's momentum distribution after releasing it from the trap~\cite{Stenger:1999, StamperKurn1999eop, Steinhauer:2002, Ozeri:2005, Richard:2003, Esslinger:2004, Papp:2008, Vale:2010, Fabbri:2011, Vale:2013,Vale:2019, Lopes:2017,Petter:2020}. Such observables are expected to be directly proportional to the DSF amplitude $S_\tau(k,\omega)$ Fourier-broadened by the finite probe time, neglecting the initial population of the high-$k$ state~\cite{Blakie:2002,Pitaevskii:2016}. By probing the system's response at a given momentum and varying $\omega$, one typically records a resonant curve of $S_\tau(k,\omega)$ with $\omega$ and the value on resonance finally gives access to $S$. This scheme has been applied to supersolids in the measurements of ref.~\cite{Petter:2020} concomitant to the present work.}

%%%%%%%%%%%%%%%%%%%%%%%%%%%%%%%%%%%%%%%%%%%%%%%%%%%%%%%%%%%%%%%%%%%%%

\section{Shallow-supersolid regime}
\label{sec:shallowssp}

In Sec.\,\ref{sec:general}, I gave general formulae expressing two characteristic properties of a SSS, namely its density contrast $C$ [Eq.\,\eqref{eq:Cnop}] and its superfluid fraction $\fsf$ [Eq.\,\eqref{eq:fsfut}],  as well as our observable of interest, the high-energy resonant scattering response $S$ [Eq.\,\eqref{eq:Sq}], as a function of the state's macroscopic field and its modulations. I remind  that, whereas the above equations give the formulae for one classical field, fluctuations effects could be accounted for by sampling the fields' phase space and performing ensemble averages of these expressions (see discussion in Sec.\,\ref{subsec:decomp}).  Based on these formulae, the aim of Secs.~\ref{sec:shallowssp} and \ref{sec:deepssp}, is to specify the scaling laws relating these three quantities, while focusing on limiting regimes, where the SSS presents extreme characters in its solid and superfluid properties. 

The present section focuses on the case of a weakly modulated SSS, i.e. with a modulated field of small amplitude compared to the uniform one. In this weak-modulation regime, here referred to as a \emph{"shallow"} supersolid, one expects $C\ll 1$ and $\fsf \approx 1$ [Eqs.~\eqref{eq:Cnop}-\eqref{eq:fsfut}]. Such a regime is of particular interest to describe the transition region between a uniform superfluid and a supersolid state, provided that the discontinuity at the transition is not too large~\cite{Blakie:2020,Sepulveda:2008}. In experiments, the dynamical crossing of this transition is one of the usual paths towards SSSs~\cite{Tanzi:2019,Boettcher:2019,Chomaz:2019}, making the present regime of practical relevance.

\begin{figure}[ht]
	\includegraphics[width=\columnwidth]{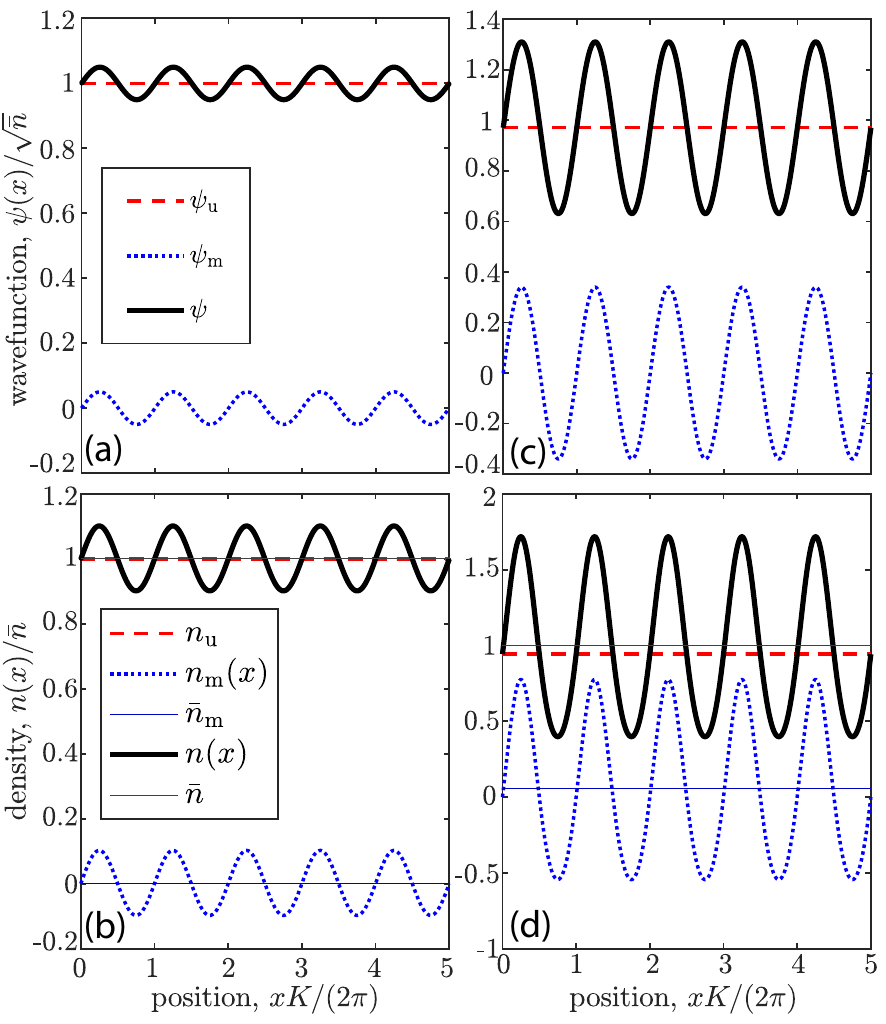}
	\caption { Examples of SSSs with sine-modulated fields, defined by Eqs.\,\eqref{eq:decpsi} and \eqref{eq:psioSIA}, using $C=0.1$ (a,b) and $C=0.7$ (c,d). Here $LK/2\pi=5$ is chosen.
	(a,c) show the axial field $\psit(x)$ (solid black line), which is here real, and its general decomposition [Eq.\,\eqref{eq:decpsi}] in $\psiu$ [dashed red line, Eq.\,\eqref{eq:psiu}] and $\psio$ [dotted blue line]. (b,d) show the corresponding density profiles and average values. The total density, $n(x)$, [thick black line, Eq.\,\eqref{eq:nx}] has for average , $\Nt$, [thin grey line, Eq.\,\eqref{eq:nt}], which decomposes in  $\Nun$ [dashed red line] and $\No$ [thin blue line] the contributions of $\psiu$ and $\psio$, respectively. The dotted blue line shows the effective profile $\nox(x)$ [Eq.\,\eqref{eq:nop}]. For (a,b) we numerically find a density contrast $C=0.1$, a superfluid fraction $\fsf=0.995$ and a high-energy scattering response $S/\Nt L=0.99875$, matching the perturbative results of Eqs.\,\eqref{eq:fsfC}-\eqref{eq:SC}. For (c) and (d), we find $C=0.62$ ($0.7$), $\fsf=0.774$ ($0.755$) and $S/\Nt L=0.942$ ($0.939$) from numerics (perturbative formulae using $C=0.7$).
	}
	 \label{fig:figSMA} 
\end{figure}
\subsection{Scaling laws of the supersolid's properties}
\label{subsec:shallowssp_gen}

To describe the shallow-supersolid regime, I introduce the normalized modulated field $\psimb(x) = \frac{\psio(x)}{\sqrt{\Nt}}$. I then define the small parameter $\Lambda \ll1$ via 
\begin{eqnarray}\label{eq:Lambda}
\Lambda^2=\int |\psimb(x)|^2dx =\frac{\No}{\Nt},
\end{eqnarray}
such that $|\psimb|$ is of order $\Lambda$.  To characterize the SSS's properties, I first derive the spatial density distribution \eqref{eq:nx}. The modulated contribution \eqref{eq:nop} rewrites as a function of $\psimb$ and $\Lambda$ as
\begin{eqnarray}
\nox(x)%=\Nt\left(|\psimb(x)|^2+2\sqrt{1-\Lambda^2}{\rm Re}[\psimb(x)]\right)\\
\approx \Nt\left(2{\rm Re}[\psimb(x)]+ |\psimb(x)|^2 +O(|\Lambda|^3)\right). \label{eq:nopsh}
\end{eqnarray}
Starting from Eq.\,\eqref{eq:nopsh}, I perturbatively expand  
Eqs\,\eqref{eq:Cnop}, \eqref{eq:fsfut} in order of $\Lambda$ to derive approximate scaling laws  of $C$ and $\fsf$. %By then combine with Eq.\,\eqref{eq:Sq} to find .

Using the fact that the real part of $\psimb(x)$ takes positive as well as negative values over space (following from Eq.\,\eqref{eq:psio}), the small density contrast reads
\begin{eqnarray}
\label{eq:Cnopsh}
C=\max({\rm Re}[\psimb])+\max(-{\rm Re}[\psimb]) +O(|\Lambda|^2).
\end{eqnarray}
It is thus of order $\Lambda$ and one can write 
\begin{eqnarray}
\label{eq:Cnopsho}
C \sim \Lambda \ll 1.
\end{eqnarray}
Leggett's formula \eqref{eq:fsfut} for the superfluid fraction rewrites as a function of $\psimb$ and $\Lambda$ as
\begin{eqnarray}
\label{eq:fsflb}
\fsf 
&=&  \left(\int\displaylimits_0^L \frac{1}{1+2{\rm Re}[\psimb]-\Lambda^2+|\psimb|^2+O(|\Lambda|^3)}\,\frac{dx}{L}\right)^{-1}.
\end{eqnarray}
Expanding the fraction up to second order in $\Lambda$ (which is the leading non-zero order) and simplifying the integral gives
\begin{eqnarray}
\label{eq:fsflb_n}
\fsf\approx 1-4\int\displaylimits_0^L \left({\rm Re}[\psimb(x)]\right)^2\,\frac{dx}{L} \geq 
1-4\Lambda^2. 
\end{eqnarray}
 Keeping in mind that Eq.\,\eqref{eq:fsfut} correctly estimates the superfluid fraction only in the case of a steady coherent state (see discussion in Sec.~\ref{subsec:ssschar}), the physically relevant case for the inequality of Eq.\,\eqref{eq:fsflb_n} is thus the equality,  $\fsf \approx 1-4\Lambda^2=1-4\No/\Nt$, with $\psimb(x)$ being real. Generally speaking, the value $1-4\No/\Nt$ extracted from the mean densities is a lower bound for the value of $\fsf$, meaning that phase modulations in the field yield an effective increase of the quantity $\fsf$. This statement should however be pondered by reminding that the occurrence of phase modulations also changes the magnitude of the modulated field. In particular, as $\int |\psit(x)|dx\geq |\int \psit(x)dx|$, the value of $\No/\Nt$ is larger for $\psit(x)$ than for a state whose field would be $|\psit(x)|$, and $1-4\No/\Nt$ is not the superfluid fraction associated to the mere amplitude of $\psit$, see also Appendix \ref{appendix:NuNt}.

Combining Eqs.\eqref{eq:fsflb_n}, and \eqref{eq:Cnopsho} with \eqref{eq:Sq}, one finds the following approximate relations between $S$, $\fsf$, and $C$:
\begin{eqnarray}
\label{eq:Sqfsf}
\frac{S}{\Nt L} &\lesssim&  \frac{\fsf+3}{4},\\
\label{eq:approxfsfCrel}
1-\fsf &\sim& C^2, \\
\label{eq:approxSCrel}
 1-\frac{S}{\Nt L} &\sim& C^2.
\end{eqnarray}
Those approximate relations are valid at leading order for small $\Lambda$ or equivalently small $C$. In Eq.\,\eqref{eq:Sqfsf}, the approximate inequality becomes an approximate equality in the case of $\psit(x)$ real, i.e. of a fully coherent stationary state.  The Equations \eqref{eq:Sqfsf}-\eqref{eq:approxSCrel} constitute the central result of Sec.\,\ref{sec:shallowssp}, specifying how the observable $S$ from high-energy scattering scales with the density contrast and superfluid fraction of a SSS in the weakly modulated regime. The scaling is mostly linear with $\fsf$ and quadratic with $C$. In the next subsection, we will exemplify these relations.

\subsection{Sine-modulated Ansatz}
\label{subsec:shallowssp_SIA}
\begin{figure}[ht]
	\includegraphics[width=\columnwidth]{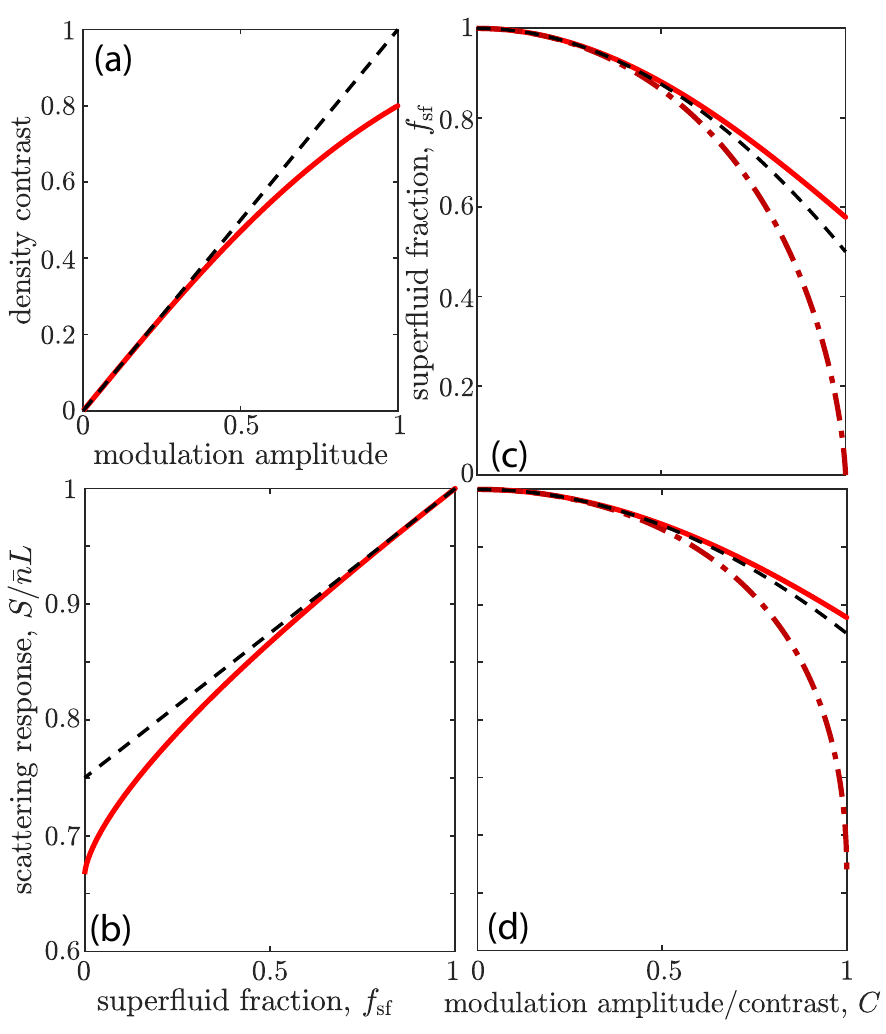}
	\caption { Illustration of the scaling laws of the SSS's properties within the sine-modulated Ansatz \eqref{eq:psioSIA} and the shallow-supersolid approximation.
	(a) density contrast numerically extracted using Eq.\,\eqref{eq:C} as a function of the amplitude modulation parameter $C$ in \eqref{eq:psioSIA} (red solid line). In the shallow supersolid regime the two quantities match (dashed line). (b) Scattering response $S/\Nt L$ as a function of the superfluid fraction $\fsf$. The solid red line shows the numerically extracted values using Eqs.\,\eqref{eq:Sq} and \eqref{eq:fsf}. The dashed line shows the expected approximate linear scaling law of Eq.\,\eqref{eq:Sqfsf}. The panels (c,d) detail the scaling law as a function of $C$ of $\fsf$ and $S/\Nt L$, respectively. The red lines show the numerically extracted values, as a function of the Ansatz's amplitude modulation [Eq.\,\eqref{eq:psioSIA}] (solid line) or as a function of the numerically extracted density contrast (dash-dotted line), see also (a). The black dashed lines show the shallow-supersolid approximate scaling laws from Eqs.\,\eqref{eq:fsfC} and \eqref{eq:SC}.    
	}
	 \label{fig:figSMA_prop} 
\end{figure}

To give a more concrete example of shallow SSS, one can use a special form of modulated field. Following the considerations of Refs.\,~\cite{Blakie:2020,Lu2015sds,Sepulveda:2008}, I apply a sine Ansatz to the modulation in the state's wave function, writing 
\begin{eqnarray}
\label{eq:psioSIA}
\psimb(x) = \psio (x)/\sqrt{\Nt} = C \sin(Kx)/2.
\end{eqnarray}
Here, $C$ is the amplitude modulation parameter, $K$ is the wave number associated with the density modulation. In order to satisfy periodic boundary conditions within the Ansatz \eqref{eq:psioSIA}, 
$LK/2\pi$ should be an integer. 
Figure \ref{fig:figSMA} shows two examples of SSSs built on the Ansatz \eqref{eq:psioSIA}. The panels (a,b) exemplify  a SSS with a weak amplitude modulation of $C=0.1$ whereas the panels (c,d) show a more strongly modulated case with $C=0.7$. In the latter case, $\Nun$ visibly deviates from $\Nt$ (see Fig.\,\ref{fig:figSMA}(d)), thus departing from the assumptions of the present Section\,\ref{sec:shallowssp}. 

A simple integration of Eq.\,\eqref{eq:psioSIA} yields  
$\Lambda= C/\sqrt{8}$. Following Eq.\,\eqref{eq:Cnopsh}, the density contrast indeed matches the Ansatz's modulation parameter, $C$, at leading order in $\Lambda$, i.e.\,for small $C$ values. Satisfying this condition, the case of Fig.\,\ref{fig:figSMA}(a,b) has a density contrast and a modulation parameter that match. In contrast, in Fig.\,\ref{fig:figSMA}(c,d), the actual density contrast equals $0.62$. Figure\,\ref{fig:figSMA_prop}(a) further details how the actual density contrast scales with the Ansatz's amplitude modulation parameter over the full range $0\leq C\leq 1$. The identity of the two quantities expected in the shallow supersolid regime is observed for a wide range of $C\lesssim 0.5$. For larger $C$, the density contrast is reduced compared to the amplitude parameter.

The sine-modulated Ansatz \eqref{eq:psioSIA} additionally enables to confirm and specify the approximate quadratic scaling laws derived in Eqs.\,\eqref{eq:approxfsfCrel}-\eqref{eq:approxSCrel} of the superfluid fraction and scattering response versus density contrast. Inserting the expression of $\Lambda$ as a function of $C$ in Eqs.\,\eqref{eq:fsflb_n} and \eqref{eq:Sq} yields
\begin{eqnarray}
\label{eq:fsfC}
\fsf\approx 1-C^2/2,\\
\label{eq:SC}
S \approx (1-C^2/8)\,\Nt L,
\end{eqnarray}
using that $\psit(x)$ is here real. Figure \ref{fig:figSMA_prop}(b-d) illustrates the relative scaling laws of $\fsf$, $S$ and $C$  within this sine-modulation Ansatz and over the full range of modulation parameter values $0\leq C \leq 1$. As for the contrast's scaling law discussed above, see Fig.\,\ref{fig:figSMA_prop}(a), the shallow-supersolid predictions are found to describe well the numerically extracted values over a relatively broad parameter range. Figure\,\ref{fig:figSMA_prop}(b) reports on the relative behavior of $S$ and $\fsf$. The approximate linear scaling law of Eq.\,\eqref{eq:Sqfsf}, generally expected in the shallow-supersolid regime, appears here relevant for $\fsf \gtrsim 0.6$.  Figure\,\ref{fig:figSMA_prop}(c,d) show that the expected quadratic scaling laws of $\fsf$ and  $S$ as a function of $C$, given by Eqs.\,\eqref{eq:fsfC}-\eqref{eq:SC}, are well satisfied for density contrasts, $C\lesssim 0.5$.  Interestingly, the shallow-supersolid scaling laws still describe well the properties' variations at larger $C$, yet when expressed in terms of the modulation parameter (solid line) and not of the actual density contrast (dash-dotted line).

%%%%%%%%%%%%%%%%%%%%%%%%%%%%%%%%%%%%%%%%%%%%%%%%%%%%%%%%%%%%%%%%%%%%%%%%%%%%%%%%%%%%%
\section{Deep-supersolid regime}
\label{sec:deepssp}

This section explores the opposite limiting case of a deep supersolid, i.e.\,a state with a strong amplitude modulation such that $C \approx 1$. In this case, as the spatial dependence of the field is in the dominant term, perturbative treatments from the general decomposition of Eqs.\,\eqref{eq:decpsi}-\eqref{eq:psio} do not easily help to simplify the expression of $\fsf$ and $C$, and to ultimately derive their relationships with $S$. 
To go further, I therefore rely on a specific Ansatz for the deep SSS's field.

\subsection{The drop-array Ansatz}
\label{subsec:DAA}
In order to describe a strongly modulated SSS, a standard model, previously used in the context of dipolar gases, assumes that the field $\psit(x)$ consists of a set of $N_D$ widely spaced and \emph{localized} density peaks (or drops)~\cite{Wenzel:2017,Chomaz:2019,Blakie:2020}. 
The present work considers such a model, additionally assuming the drops to be regularly spaced by a distance $d$ along $x$ and to have identical density profiles. This implies that the drops can be described by a unique wave function, $\chi(x)$ here taken to be centered at $x=0$. Further, $|\chi(x)|$ is assumed to be symmetric and monotonically decreasing towards zero for increasing $|x|$ (c.f.\,"localized" peak). A \emph{localization} parameter, $\lambda$, is defined by
 the ratio of the root-mean-square (rms) width of the drop density profile to the distance $d$,  i.e.\,$\lambda = \sqrt{\int x^2|\chi(x)|^2 dx/N_{1d}}\big/d$ where $N_{1d}= \int |\chi(x)|^2 dx$ is the number of atoms in one drop.

In our model, the drops can additionally have independent phase profiles. As introduced in Sec.\,\ref{subsec:decomp}, the inclusion of phase patterns enables us to encompass various physical phenomena, ranging from dynamics to fluctuations. Of particular relevance for the deep-supersolid regime, we note that the strength of statistical phase fluctuations connects to the degree of global phase coherence of the underlying state, larger fluctuations meaning weaker coherence. In the present modeling, the phases are assumed to be uniform over each drop, with value $\theta_j$ for the drop $j$. This description of the phase profile over an array is reminiscent of an array of Josephson junctions~\cite{fazio:2001}. Josephson-junction models were also previously fruitfully applied to the case of density modulated states of dipolar gases, see Refs.\,\cite{Wenzel:2017,Chomaz:2019,Ilzhofer:2019}. The phase-coherent stationary case where $\psit(x)$ is real corresponds to $\theta_j \equiv 0$. 

Finally,  the total field for the drop array writes
\begin{eqnarray}
\label{eq:dropphi}
\psit(x) &= \sum_{j=0}^{N_D} \chi(x-jd)e^{i\theta_j}.
\end{eqnarray}
In order to satisfy periodic boundary conditions within the Ansatz \eqref{eq:dropphi}, $L$ should be $L=d(N_D-1)$ and $\theta_1=\theta_{N_D}$. There is then effectively only $N_D-1$ independent drops in the array.

For the Ansatz \eqref{eq:dropphi} to be relevant (i.e.\,to  be able to indeed isolate drops), the wave-function overlap between neighboring drops is required to be small. This requirement imposes a steep-enough functional dependence of $|\chi(x)|$ over the range set by the drop distance $d$. Relying on the monotonic character of $|\chi(x>0)|$, we generally define the steep-$|\chi|$ condition as 
\begin{eqnarray}
\label{eq:steep_chi}
\frac{|\chi\left(d\right)|}{|\chi\left(d/2\right)|}\ll \frac{|\chi\left(d/2\right)|}{|\chi\left(0\right)|}\ll 1,
\end{eqnarray}
see also Sec.\,\ref{subsec:DAA_dens} and Appendix \ref{appendix:ndrop}. The steep-$|\chi|$ condition \eqref{eq:steep_chi} has strong connections to a \emph{localization} condition, constraining the localization parameter (i.e.\,drops' extent-over-distance ratio) to small values, $\lambda\ll 1$. In particular, $|\chi(d/2)| \sim |\chi(0)|$ generally implies $\lambda\sim 1$, and, for a fixed functional form for $|\chi(x)|$, the ratios appearing in Eq.\,\eqref{eq:steep_chi} decrease with $\lambda$.

In the present model, one should note that, if the drops' phases $\theta_j$ are not equal, then the wave function's phase would jump in between the drops. In a more realistic model, the phase gradients should be finite. Due to the continuity equation, it is however physically expected that, in a steady state, phase gradients are concentrated in the region of lower densities~\cite{Leggett,Pitaevskii:2016}. In a array of steep drops (see also Sec.\,\ref{subsec:DAA_dens}), this is exactly in between the drops, as assumed by the present model.

 In the following, most conclusions will be drawn without specifying a functional form for $\chi(x)$ and simply assuming it to be steep [Eq.\,\eqref{eq:steep_chi}].
In the context of dipolar gases, drop Ansätze were found to be appropriate in  regimes similar to that of the present work, describing either single-drop states or density-modulated ones, see Refs.\,\cite{Blakie:2020,Wenzel:2017,Chomaz:2019,Waechtler:2016b,Baillie:2018}. These works typically rely on a specific Ansatz for $\chi(x)$, assuming it of Gaussian profile. In the following, I will use such a Gaussian Ansatz to exemplify the relationships generally derived, i.e. posing 
\begin{eqnarray}
    \label{eq:chig}
    \chi(x)= \chi_{\rm g}(x) \equiv \sqrt{\frac{N_{1d}}{\pi^{1/2}\lambda d}}\exp\left(-\frac{x^2}{2\lambda^2 d^2}\right). 
\end{eqnarray}

\begin{figure}[ht]
	\includegraphics[width=\columnwidth]{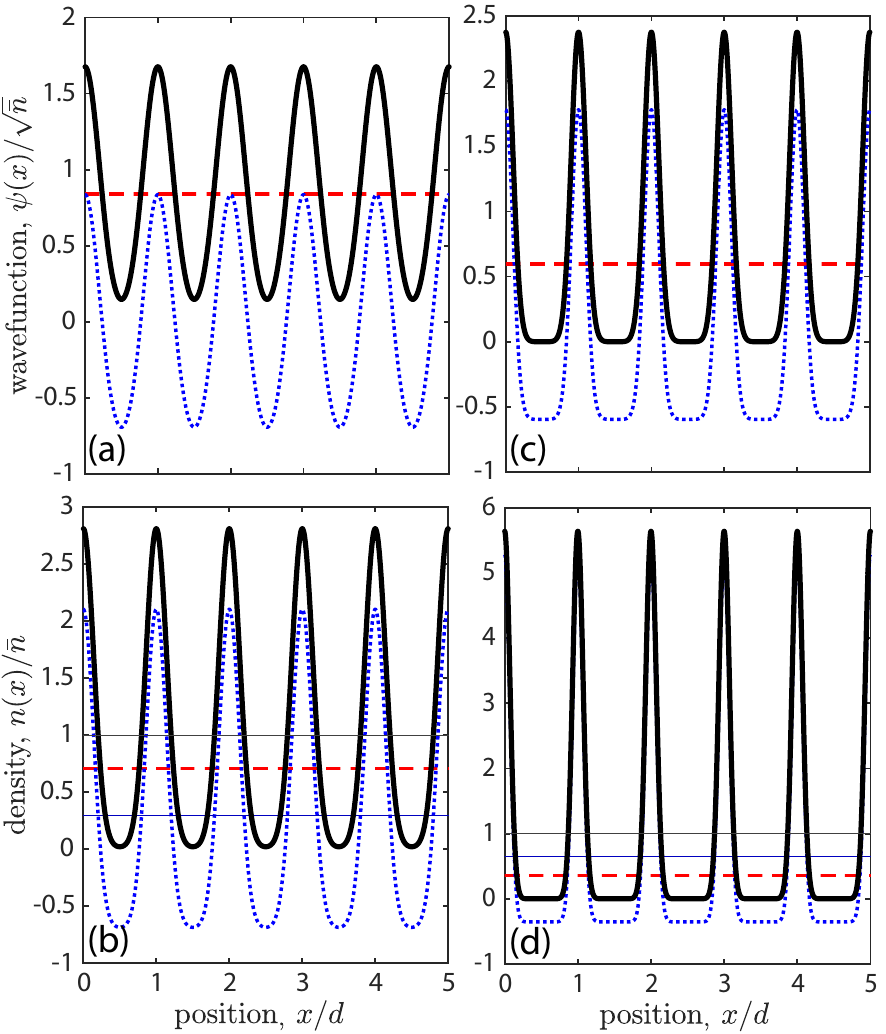}
	\caption { Examples of SSSs based on the drop-array Ansatz, defined by Eqs.\,\eqref{eq:dropphi}, and \eqref{eq:chig}, using $\lambda=0.2$ (a,b) and $\lambda=0.1$  (c,d), and a phase-coherent stationary configuration $\theta_j= 0$. Here $L= 5d$ is chosen. The panels and the color code are similar to those of Fig.\,\ref{fig:figSMA}.
	(a,c) show $\psi(x)$ (solid black line), and its general decomposition in $\psiu$ (dashed red line) and $\psio$ (dotted blue line). (b,d) show the corresponding density profiles and average values: the total density $n(x)$, (thick black line), its average $\Nt$ (thin grey line), the uniform contribution $\Nun$ (dashed red line), the average modulated contribution $\No$ (thin blue line), and the effective profile $\nox$ (dotted blue line).  For a,b (c,d) we find $1-C=0.015$ ($1.1 \cdot 10^{-10}$), $\fsf=0.11$  ($7.6 \cdot 10^{-9}$) and $S/\Nt L=0.71$ ($0.35$) both from numerics and from the perturbative formulae, Eqs.\,\eqref{eq:CdeepG},\eqref{eq:fsfCg}-\eqref{eq:Sq_chi}).
	}
	 \label{fig:figDAA} 
\end{figure}

Examples of phase-coherent stationary Gaussian-drop arrays are shown in Fig.\,\ref{fig:figDAA} for two values of $\lambda$. The panels (a,b) show an array made of relatively broad drops with $\lambda=0.2$, and (c,d) exemplify an array of much steeper drops with $\lambda=0.1$. The upper row shows the states' wave function following from Eqs.\,\eqref{eq:dropphi} and \eqref{eq:chig}.  
 The lower panels show the corresponding density profiles, which will be the focus of the next Section \ref{subsec:DAA_dens}.

%%%%%%%%%%
\subsection{Spatial density in the drop array}
\label{subsec:DAA_dens}
To express our physical observables $C$, $\fsf$, and $S$ for the drop-array model \eqref{eq:dropphi}, we start, as in the previous sections, by writing the density profile. It generally is
\begin{eqnarray}
\label{eq:nxdrop_sumjl}
n(x) &=&  \sum_{j,l=0}^{N_D-1} e^{i(\theta_j-\theta_l)} \chi(x-jd)\chi^*(x-ld).
\end{eqnarray}
The steep-$|\chi|$ condition \eqref{eq:steep_chi} allows to simplify Eq.\,\eqref{eq:nxdrop_sumjl} by sorting the terms by magnitude orders for each position of space, see also  Appendix \ref{appendix:ndrop} for further details. 

The terms of the sum of  Eq.\,\eqref{eq:nxdrop_sumjl} correspond to the  wave-function overlaps for the drop couple $(j,l)$ . These terms can be relevantly differentiated as a function of the drops' spacing, $s=l-j$. Equation \eqref{eq:nxdrop_sumjl} then rewrites
\begin{eqnarray}
\label{eq:nxdrop_sumjl_s}
n(x) &=&  \sum_{j=0}^{N_D-1} \sum_{s=-j}^{N_D-j-1} e^{i(\theta_j-\theta_{j+s})} \eta_s(x).
\end{eqnarray} 
where $\eta_s$ is the general $s$-overlap function, not accounting for the drop's individual phases but for their identical shapes, and defined via
\begin{eqnarray}
    \label{eq:etadef}
    \eta_s(x)=\chi(x)\chi^*\left(x-sd\right). 
\end{eqnarray}
The $\eta_{0}$-function matches the individual drop density profile, maximum at $x=0$. The $\eta_{1}$-function is the nearest-neighbor overlap. Under the  steep-$|\chi|$ assumption \eqref{eq:steep_chi}, $\eta_{1}$ is maximum at  $x\approx d/2$, with a value $\eta_{1}(d/2)\approx \eta_0(d/2)$. Under this same approximation, the $\eta_s$-functions are steeply decreasing in magnitude with $s$. Therefore, 
at any $x$, in the sum of Eq.\,\eqref{eq:nxdrop_sumjl_s}, one can neglect the $s>$1-contributions, giving 
\begin{eqnarray}
n(x)\approx \ndrt(x) + \nov(x).
\label{eq:nxd_sumjl_split}
\end{eqnarray}
The $s$=\,0-contribution, $\ndrt(x)$, matches the density of all the individual drops,
\begin{eqnarray}
    \label{eq:ndrt}
    \ndrt(x)=\sum_{j=1}^{N_D} \eta_0(x-jd).
\end{eqnarray}
It dominates overall. As the steep-$|\chi|$ condition \eqref{eq:steep_chi}  additionally implies that, at the drop $j$'s center, the contributions of all other drops can be neglected, $\ndrt(x)$ has maxima at each $x \approx jd$, approximately equaling $\eta_0(0)$, and minima at $x\approx (j+1/2)d$, approximately equaling $2\eta_0(d/2)$, from the contributions of the two neighboring drops. 
The $s$=\,1-overlap contribution, $\nov(x)$, writes
\begin{eqnarray}
    \label{eq:nov}
    \nov(x)=2 \sum_{j=1}^{N_D-1} {\rm Re}\left[e^{i\Delta \theta_j}\eta_{1}(x-jd)\right].
\end{eqnarray}
with $\Delta \theta_j= \theta_j-\theta_{j+1}$. The contribution $\nov(x)$ competes with $\ndrt(x)$ only at $x\approx (j+1/2)d$. In absence of phase profiles (phase-coherent stationary case), the two contributions have the same magnitude at these intermediate positions, $\ndrt((j+1/2)d) \approx \nov((j+1/2)d) \approx 2\eta_0(d/2)$. The complex numbers $e^{i\Delta \theta_j}$ coming into play in Eq.\eqref{eq:nov} tend to make the overlap contribution reduce, and even counteract (by changing sign) the presence of  $\ndrt(x)$, in between the neighboring drops ($j$,$j+1$) when their phase difference increases. When the two drops are $\pi$-shifted, the total density $n(x)$ between them approximately cancels out as $\nov((j+1/2)d) \approx -\ndrt((j+1/2)d) \approx -2\eta_0(d/2)$.  In presence of statistical phase fluctuations, this means that the total density $n(x)$ in between the drops reduces with the degree of phase coherence. 

By integrating Eq.\,\eqref{eq:nxd_sumjl_split} over $x\in [0,L]$, one gets the mean density $\Nt$. At leading perturbation order, it is dominated by the $s$=0 contribution and, using periodic boundary conditions (see discussion below Eq.\,\eqref{eq:dropphi}, $L=(N_D-1)d$ and the array effectively contains $N_D-1$ drops), it writes
\begin{eqnarray}
\label{eq:ntdrop}
\Nt&\approx&  \frac{(N_D-1)N_{1d}}{L} =  \frac{N_{1d}}{d}.
\end{eqnarray}

The Gaussian-drop model \eqref{eq:chig} enables to exemplify the above general derivations. In this case,
\begin{eqnarray}
    \label{eq:etadefg}
&\eta_s(x) =&\eta_0(0)\exp\left(\frac{-(x/d)^2-(x/d-s)^2}{2\lambda^2}\right)\\
\label{eq:etadefg0}
&\text{with } \eta_0(0) &= \Nt/\sqrt{\pi}\lambda.
\end{eqnarray}
 The steep-$|\chi|$ condition \eqref{eq:steep_chi} here simply writes $\exp(-1/4\lambda^2)\ll 1$. In the Gaussian-drop Ansatz, the drop's extent-over-distance ratio $\lambda$ is thus the parameter controlling the drop's steepness and the steep-drop condition appears to be satisfied once $\lambda\lesssim 0.3$. 
 Examples of Gaussian-drop-array density profiles are given in the lower panels of Fig.\,\ref{fig:figDAA}. Matching the above description, the density profiles show maxima at the drop positions and minima in between. This is true both for the relatively broad drops (b) and for steeper ones (d). The density in between the drops sharply decreases from  $\lambda=0.2$ to  $\lambda=0.1$, as expected from the $\exp(-1/4\lambda^2)$-scaling law of $\eta_0(d/2)/\eta_0(0)$.

\subsection{Density contrast in the drop array}

Summarizing the above discussion, under the steep-$|\chi|$ assumption \eqref{eq:steep_chi}, 
the total density, given by Eq.\,\eqref{eq:nxd_sumjl_split}, has maxima at $x\approx jd$, approximately equaling $n(jd) \approx \eta_{0}(0)$, and local minima at $x \approx (j+1/2)d$, approximately equaling $n((j+1/2)d) \approx 2(1+\cos \Delta \theta_j)\eta_{0}(d/2) 
\ll n(jd)$. A first consequence of the above discussion is that, at leading order, the density profile is nearly fully contrasted, $C\approx 1$; see also Fig.\,\ref{fig:figDAA}(b,d). This qualifies the drop-array state as a deep supersolid under the steep-$|\chi| $ condition \eqref{eq:steep_chi}, justifying the use of Eqs.\eqref{eq:chig}-\eqref{eq:phi} in the remainder of Sec.\,\ref{sec:deepssp}.

At next order, we should specify the overall minimum. This is $2\phi\eta_{0}(d/2)$ where 
\begin{eqnarray}
\label{eq:phi}
\phi=1+\min_j \left[\cos \Delta \theta_j\right]
\end{eqnarray} 
is a numerical factor, ranging between 0 and 2, and characterizing the phase variations on the drop array. For a phase-locked array, $\theta_j\equiv 0$, as relevant for a fully coherent steady state, $\phi=2$. For a strongly phase-modulated array such that the phase of a drop can be $\pi$ shifted compared to its neighbor, $\phi=0$. For what concerns fluctuations, as highlighted above, the strength of the fluctuations in the drops' phases relates to the degree of coherence that thus connects to a reduced $\phi$. 

 The density contrast then reads
\begin{eqnarray}
\label{eq:Cdeep}
C \approx 1-4\phi\frac{\eta_{0}(d/2)}{\eta_{0}(0)}.
\end{eqnarray}
It is reduced compared to the full-contrast value by two features of the array, (i) the relative value of the individual drop's density remaining at half-distance from its neighboring drop, compared to its value at center, (ii) the degree of phase locking in the array, as encompassed by $\phi$. 
We note that the degree of phase locking coming into play in the contrast value is a rather local one, as $\phi$ is sensitive to phase differences in between neighboring drops only.

\begin{figure}[ht]
	\includegraphics[width=\columnwidth]{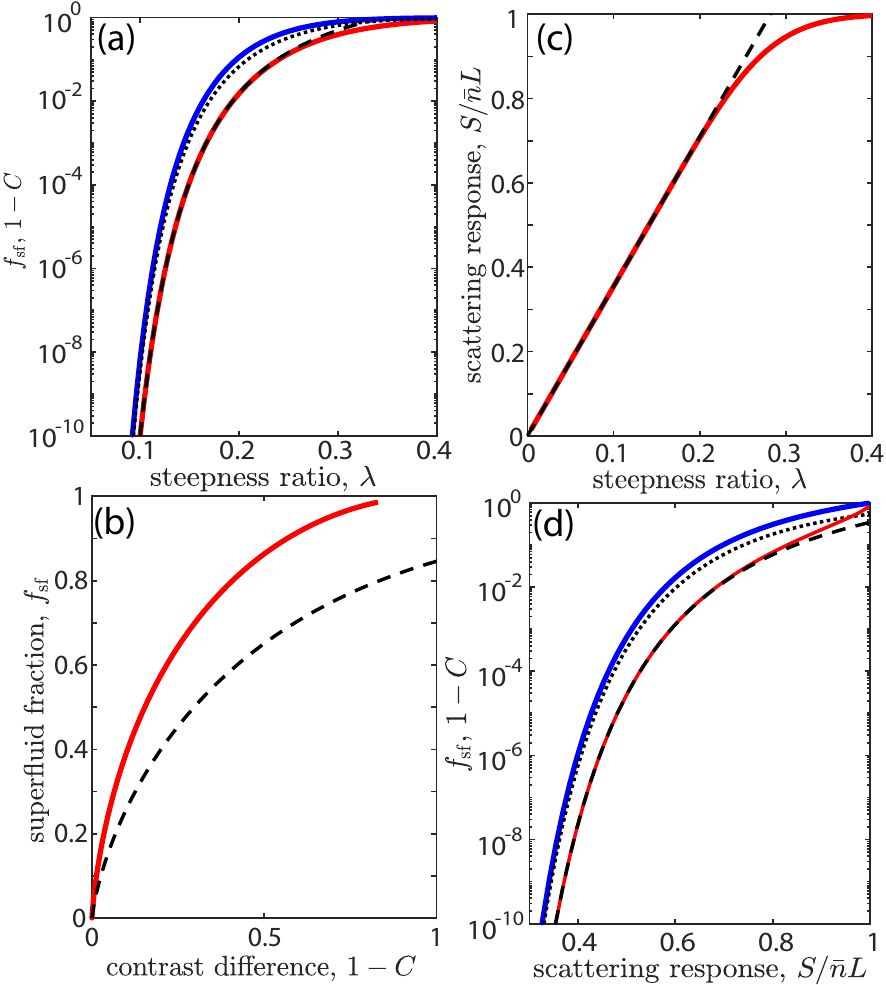}
	\caption { Illustration of the scaling laws of the SSS's properties within the Gaussian drop-array Ansatz [Eqs.\,\eqref{eq:chig},\eqref{eq:dropphi}] and deep-supersolid approximation. The array is here assumed fully phase coherent.
	(a) remainder of the density contrast $1-C$ and superfluid fraction as a function of the parameter controlling the drop steepness, $\lambda$. Numerically extracted values from the full density profiles are represented by the solid lines, red and blue, respectively. Approximate scaling laws from the steep-$|\chi|$ condition (Eqs.\eqref{eq:CdeepG} and \eqref{eq:fsfCg}) are given by the black lines, dashed and dotted, respectively. (b) $\fsf$ versus $1-C$ from numerical extractions (solid red line) and approximate scaling laws (dashed line). (c) scattering response $S/\Nt L$ as a function of $\lambda$ from a numerical extraction (solid red line) and approximate scaling law of Eq.\,\eqref{eq:Sq_chi} (dashed line). (d) scaling law of $\fsf$ versus $1-C$ as a function of $S/\Nt L$, same colorcode as (a).   
	}
	 \label{fig:figDAA_prop} 
\end{figure}

In the Gaussian-model case \eqref{eq:chig}, the contrast reads 
\begin{eqnarray}
\label{eq:CdeepG}
C \approx 1-4\phi\exp\left(-\frac{1}{4\lambda^2}\right).
\end{eqnarray}
and its difference to unity steeply decreases with decreasing $\lambda$. This is exemplified in Fig.\,\ref{fig:figDAA} where $1-C$ is reduced by eight orders of magnitude from $\lambda=0.2$ to $\lambda=0.1$. The steep scaling law of the contrast with $\lambda$ for arrays of Gaussian drops is further illustrated in the semi-logarithmic plot of Fig.\,\ref{fig:figDAA_prop}(a). The approximate scaling law of Eq.\,\eqref{eq:CdeepG} is observed up to $\lambda\lesssim 0.3$.

\subsection{Superfluid fraction within the drop-array Ansatz}
\label{subsec:DAA_fsf}

To evaluate the superfluid density using Leggett's formula [Eq.\,\eqref{eq:fsf}], one has to consider the integral of the inverse density profile. Starting from Eqs.\,\eqref{eq:nxd_sumjl_split}-\eqref{eq:nov}, the terms of the sums that effectively contribute to $n(x)$ at a given $x$, depend on $x$. As under the steep-$|\chi|$ condition \eqref{eq:steep_chi}, a drop does not contribute significantly to the density profile at distance larger than $d$ of its center, one can decompose the integral of Eq.\,\eqref{eq:fsf} by segment $x\in[jd,(j+1/2)d]$. Also using $L=(N_D-1)d$, Eq.\,\eqref{eq:fsf} writes 
\begin{eqnarray}
\nonumber
&&\fsf \approx  \left(\frac{\Nt}{d(N_D-1)} \sum_{j=0}^{N_D-1} \sum_{s=-1}^{0}\right.\\
\label{eq:fsfeta}
&&\left.\int\displaylimits_0^{d/2}\frac{dx}{\eta_{0}(x)+ \eta_{0}(x-d) 
+ 2{\rm Re}\left[e^{i \Delta \theta_{j+s}}\eta_{1}(x)\right]}\right)^{-1}.
\label{eq:fsfeta2}
\end{eqnarray} 
Equation \eqref{eq:fsfeta} includes the phase factors $e^{i \Delta \theta_j}$. Yet it should be kept in mind that, as discussed above (Sec.\,\ref{subsec:ssschar}),  $\fsf$ is physically meaningful only for a phase-coherent stationary array, i.e.\,$e^{i \Delta \theta_j}\equiv1$. 

The $x$-range dominating the integral of Eq.\,\eqref{eq:fsfeta} is where the density is the smallest, i.e.\,at $x\approx d/2$. In this range, the denominator approximately equals $2(1+\cos\Delta\theta_j)\eta_0(d/2)$. The spatial extent of the low-density region depends on the steepness of $|\chi|^2$ in this $x$-range. A rough estimate, setting this extent to the rms width $\lambda d$ and $\Nt \sim \eta_0(0)\lambda$ [see Eq.\,\eqref{eq:etadefg0}], gives
\begin{eqnarray}
\label{eq:fsfetasim}
\fsf \sim  \frac{\eta_0(d/2)}{\eta_0(0)}\frac{1}{\lambda^2} \phi',
\end{eqnarray}
with 
\begin{eqnarray}
\label{eq:phip}
\phi'=1+\langle\cos\Delta\theta_j\rangle_j \geq \phi,
\end{eqnarray}
being a numerical phase factor similar to $\phi$ [Eq.\eqref{eq:phi}], but here using the average over the array, $\langle.\rangle_j$, instead of the minimum. Similar to $\phi$, the factor $\phi'$ probes the local phase modulations on the array, i.e.\,at the level of neighboring drops. In the physically relevant phase-coherent stationary case, $\phi'\equiv 2$ and this factor can be discarded from the scaling law \eqref{eq:fsfetasim}.

The smallness of $\fsf$ depends on the steepness of the drop's wave function: $\fsf$ is small only if  $|\chi(x)|$ decreases faster than $1/x$ for large $x \gtrsim d/2$. Using Eq.\,\eqref{eq:Cdeep}, one also finds an approximate scaling law with the density contrast, 
\begin{eqnarray}
\label{eq:fsfCsim}
\fsf \sim  \frac{\phi'}{\phi}\frac{1-C}{\lambda^2}.
\end{eqnarray}

Using a Gaussian-drop model \eqref{eq:chig} enables us to go beyond this rough estimate. Here Eq.\,\eqref{eq:fsfeta} yields
\begin{eqnarray}
\label{eq:fsfetag}
\fsf &\approx&  \frac{\phi'\exp\left(-\frac{1}{4\lambda^2}\right)}{\lambda^2}\mathcal{I}^{-1},\\
\mathcal{I} &=& \int_{-\frac{1}{2\lambda}}^{\frac{1}{2\lambda}} \frac{  \sqrt{\pi}\exp(u^2)du}{(\phi-1)+ \cosh\left(\frac{u}{\lambda}\right)}.
\end{eqnarray}
In the fraction of Eq.\,\eqref{eq:fsfetag}, one recognizes Eq.\,\eqref{eq:fsfetasim}. 
The inverse of the integral $\mathcal{I}$ gives the numerical prefactor. For $\lambda\ll1$, $\mathcal{I}$ is well approximated by $\mathcal{I} \approx 2\pi\exp\left(-\frac{1}{4\lambda^2}\right){\rm Erfi}\left(\frac{1}{2\lambda}\right)$ and, at leading order in $\lambda$, $\mathcal{I} \approx 4\sqrt{\pi}\lambda$. Here ${\rm Erfi}$ is the imaginary error function. Therefore, $\fsf$ rewrites
\begin{eqnarray}
\label{eq:fsfCg}
\fsf \approx \frac{\phi'\exp\left(-\frac{1}{4\lambda^2}\right)}{4\sqrt{\pi}\lambda^3} \approx  \frac{1-C}{\lambda^3}\frac{\phi'}{16\sqrt{\pi}\phi} .
\end{eqnarray}
In this Ansatz, $|\chi|$ scales very steeply and $\fsf$ vanishes for decreasing $\lambda$, as shown in Fig.\,\ref{fig:figDAA_prop}(a), blue line. The scaling law of Eq.\,\eqref{eq:fsfCg} describes well the numerical evaluation of $\fsf$ from Leggett's formula \eqref{eq:fsf} up to $\lambda\lesssim 0.3$.  Furthermore, the scaling laws of both $1-C$ and $\fsf$ are dominated by the exponential function, suggesting that the two quantities scale together. 
Their relative scaling is illustrated in Fig.\,\ref{fig:figDAA_prop}(b). Here the approximation of Eq.\,\eqref{eq:fsfCg} holds quantitatively only for $C\rightarrow 1$, evidencing the sensitivity to the exact prefactors.

\subsection{High-energy scattering response within the drop-array Ansatz}
\label{subsec:DAA_DSF}
We finally focus on high-energy scattering as it is the observable of focus for the present paper. As shown in Sec.\,\ref{subsec:highscatt} [Eq.\,\eqref{eq:Sq}], the resonant response factor $S$ simply scales with the density of the uniform contribution from the decomposition \eqref{eq:decpsi}. For a drop array of wave function~\eqref{eq:dropphi}, the uniform field \eqref{eq:psiu} writes
\begin{eqnarray}
\label{eq:deepsspsiu}
\psiu  &=& \left(\sum_{j=1}^{N_D-1} e^{i\theta_j}\right)\int \chi(x) \frac{dx}{L} 
\end{eqnarray}
using periodic boundary conditions, which imply that the array effectively counts only $N_D-1$ independent drops and $L=(N_D-1)d$ (see discussion below Eq.\,\eqref{eq:dropphi}).  This yields the response factor
\begin{eqnarray}
\label{eq:Sq_chi}
S \approx \Nt L \lambda \frac{\left|\int \chi(u\lambda d) du \right|^2}{\int \left|\chi(u\lambda d)\right|^2du } \left|\frac{1}{N_D-1}\sum_{j=1}^{N_D-1} e^{i\theta_j}\right|^2 . 
\end{eqnarray}

From Eq.\,\eqref{eq:Sq_chi}, $S$ can be decomposed into two contributions, (i) a phase-locked contribution $S_0$ and (ii) a reduction factor due to the phase modulations in the array, $\nuph$. Precisely,  
\begin{eqnarray}
\label{eq:Sda}
S &\approx S_0\nuph, \text{ with } \\
\label{eq:S0da}
S_0 &=  
\lambda\left(\frac{\left|\int \chi(u\lambda d) du \right|^2}{\int \left|\chi(u\lambda d)\right|^2du }\right)\, \Nt L ,\\
\label{eq:finc}
\nuph &= \left|\frac{1}{N_D-1}\sum_{j=1}^{N_D-1}e^{i\theta_j}\right|^2.
\end{eqnarray}
The phase-locked contribution is the only relevant one for the case of a phase coherent stationary array, where as $\nuph$ encompasses additional effects from dynamics and fluctuations.   

On the one hand, the phase-locked contribution, $S_0$, is reduced compared to the unform value $\Nt L$ by a factor which is sensitive to the arrangement of the drops in the array. Interestingly, this dependence is mostly via an approximate linear scaling law with the localization parameter $\lambda$. 
This scaling law fundamentally differs from those of the superfluid fraction and of the density contrast, see Eqs.\,\eqref{eq:Cdeep}-\eqref{eq:CdeepG}. The latter scaling laws generally involve $\eta_0(d/2)/\eta_0(0)$, which depends on the steepness of $|\chi|$. If $|\chi(x)|$ is steeper than  $1/\sqrt{x}$ at large distances, the superfluid fraction and of the density contrast scales faster than $S_0$ with $\lambda$.

To specify these statements, we consider the Gaussian-drop model \eqref{eq:chig}. Here the scaling laws for the superfluid fraction and density contrast are dictated by the factor $\exp(-1/4\lambda^2)$ [see Eqs.\,\eqref{eq:fsfCsim}-\eqref{eq:fsfCg}], which is much steeper than the linear scaling of $S_0$ in $\lambda$. The contribution $S_0$ can also be specified in this Gaussian case, 
\begin{eqnarray}
\label{eq:S0dag}
S_0 =2\sqrt{\pi}\lambda \,\Nt L .
\end{eqnarray}
confirming the linear scaling of Eq.\,\eqref{eq:S0da}. As illustrated in Fig.\,\ref{fig:figDAA_prop}(c), this scaling for $S$ in the phase-locked case appears to hold up to $\lambda \lesssim 0.25$. The much weaker scaling law of $S_0$ with $\lambda$ compared to those of $1-C$ and $1-\fsf$ is further illustrated by their relative plots in semi-logarithmic scale given in Fig.\,\ref{fig:figDAA_prop}(d). This is also visible in Fig.\,\ref{fig:figDAA}, where $\Nun$, matching $S_0/L$, remains significant in highly contrasted drop arrays.

On the other hand, the phase-dependent contribution, $\nuph$, brings an additional reduction factor which is sensitive to the phase modulation in the array. Generally speaking $\nuph\leq 1$, and $\nuph=1$ if and only if the array is in a fully coherent stationary configuration (phase locked).  The comparison of Eq.\,\eqref{eq:finc} with Eqs.\,\eqref{eq:phi}-\eqref{eq:Cdeep} shows that the phase information contained in the scattering response is more global than that contained in the other quantities considered up to now. In $\phi$ ($\phi'$), which sets the sensitivity of $C$ ($\fsf$) to the phase modulation in the array, only the phase difference between neighboring drops are considered. In contrast, $\nuph$ measures the actual length of the full phase vector over the drop array in comparison to the fully stretched configuration.  By expanding the norm, Eq.\,\eqref{eq:finc} rewrites
\begin{eqnarray}
\label{eq:finc2}
\nuph &= \frac{1}{(N_D-1)^2}\sum_{j,l=1}^{N_D-1}\cos(\theta_j-\theta_l),
\end{eqnarray}
where one sees that the phase differences between all drop couples come into play regardless of their distances. As highlighted above (see e.g.\,Secs.\,\ref{subsec:decomp}, \ref{subsec:DAA}), the phase differences between drops can be deterministic, e.g. when arising from coherent dynamics, or of statistical character when connecting to thermal or quantum fluctuations of the state. 
The high-energy scattering response thus appears as a highly sensitive probe of the phase fluctuations and deterministic modulations on the full extent of a strongly density-modulated state.

\section{Conclusion}
\label{sec:conclusions}

By describing supersolid states using a classical-field approach, I derived relationships between two of their fundamental properties, namely the density contrast  and the superfluid fraction, and their resonant response to a high-energy scattering probe. I specified these relationships in two opposite regimes. 

In the case of a weakly modulated  (\emph{"shallow"}) supersolid, the high-energy scattering response generally scales approximately  linearly with the superfluid fraction [Eq.\,\eqref{eq:Sqfsf}] and quadratically with the density contrast [Eq.\,\eqref{eq:approxSCrel}] . This establishes high-energy scattering spectroscopy as a sensitive probe of the spontaneous appearance of density modulation in a superfluid. For a fully coherent stationary state and as long as $\fsf$ remains close to unity, the scattering response is additionally expected to provide an estimate of the superfluid fraction. 

In the case of a strongly modulated (\emph{"deep"}) supersolid, a drop-array Ansatz is used. In this case, the scaling laws of the physical properties depend on the steepness of the drop's wave function over the inter-drop distance. Assuming Gaussian-shaped drops, the density contrast and superfluid fraction vary very steeply with the ratio $\lambda$ of the drops' extent to their distance, mostly as $\exp(-1/4\lambda^2)$ [Eqs.\,\eqref{eq:CdeepG}, \eqref{eq:fsfCg}]. In contrast, the scattering response varies much more smoothly with this ratio, roughly linearly [Eqs.\,\eqref{eq:S0da}, \eqref{eq:S0dag}]. Hence, in contrast to the shallow-supersolid case, the sensitivity of the scattering probe to density modulation or superfluid fraction appears to be lost in the deep-supersolid regime. 
Yet, a distinct feature of the scattering response is that it has an acute sensitivity to the global modulations of the phase of the deep supersolid [Eq.\,\eqref{eq:finc}]. This makes high-energy scattering spectroscopy a promising path to investigate both coherent dynamics and fluctuation effects in supersolid systems.  

In our complementary work of Ref.\,\cite{Petter:2020}, we report on a first study of dipolar gases across the transitions from regular superfluid to supersolid and to incoherent crystal based on a high-energy two-photon Bragg-scattering technique. The simple models developed in the present work are there applied and compared to both eMF-theory and experimental results. They are found to provide an intuitive understanding of the physics at play, and to show a quantitative predictive power. They are crucial to not only understand but also demonstrate the central role played by the phase effects induced by the dynamical crossing of the superfluid-to-supersolid phase transition in experiments.

By figuring out and disentangling the sensitivity of the high-energy scattering response to both density and phase modulations occurring in self-modulated quantum states, the present work ultimately establishes high-energy scattering spectroscopy as an exquisite probe of the characteristic properties of supersolid states. {High-energy scattering is readily achievable in a variety of systems~\cite{Palevsky:1957,Henshaw:1958,Yarnell:1959,Taylor1991:DIS, Kendall:1991:DIS, Friedman1991:DIS, Breidenbach:1969,Griffin:1993, GIACOVAZZO:1992} and, in particular, in ultracold quantum gases, via two photon Bragg scattering~\cite{Richard:2003,Esslinger:2004,Papp:2008, Vale:2010, Fabbri:2011, Vale:2013,Vale:2019, Lopes:2017,Petter:2020}. This makes our study relevant for the dipolar supersolids and beyond}. Furthermore, by deriving general and simple relations, the present work provides a hopefully useful framework for future studies of such intriguing states{, in any system, as long as they can be pictured within a classical-field theory.}

\begin{acknowledgments}
 
I thank Francesca Ferlaino, Daniel Petter, Danny Bailly, Blair Blakie, Mikhail Baranov, Alessio Recati, Gabriele Natale, Alexander Patscheider, and Manfred J. Mark for stimulating discussions and comments on the manuscript. I thank especially Blair Blakie and Mikhail Baranov for many insightful explanations. I thank Danny Baillie for a careful check of my derivations. I thank Philippe Chomaz for enriching comments. This work is financially supported through a DFG/FWF grant (FOR 2247/PI2790), a joint RSF/FWF grant (I 4426), and the FWF Elise Richter Fellowship number V792.
\end{acknowledgments}

* Correspondence and requests for materials
should be addressed to lauriane.chomaz@uibk.ac.at.

\appendix

\section{Respective roles of amplitude and phase modulations in the field decomposition.}
\label{appendix:NuNt}  
The present work describes a SSS via a single macroscopic field with both amplitude and phase modulations,  see discussion in Sec.\,\ref{subsec:decomp}. Equations.\,\eqref{eq:decpsi}-\eqref{eq:nop} specify a decomposition of the field into uniform and modulated components, $\psit(x)=\psiu+\psio(x)$, which also yield a simple decomposition of the mean density $\Nt=\Nun+\No$. In this appendix, I illustrate the respective roles of the amplitude and phase modulations in these decompositions.

First, we show that any source of modulation yields a non-zero $\psio$. Indeed, using the Cauchy-Schwarz equality condition for the inequality $|\int_0^L \psit(x)dx|^2\leq \int_0^L dx \cdot \int_0^L |\psit(x)|^2dx$, one finds that $\Nun=\Nt$ only if $\psit$ is colinear with a constant function over space, i.e. if $\psit$ is uniform. This means that once a modulation, of any kind, is present, the density in the uniform component is reduced compared to $\Nt$. Below, I detail the respective effects of amplitude and phase modulations in this reduction. 

The effects of amplitude modulations can be seen from the Cauchy-Schwarz inequality $|\int_0^L |\psit(x)|dx|^2\leq L \int_0^L |\psit(x)|^2dx$. As the right-hand side matches $\Nt L^2$, and the left-hand side $\Nun L^2$ while only considering the amplitude of $\psit$, it  implies that the presence of amplitude modulations indeed results in a decrease of
 $\Nun$ with respect to $\Nt$.  
 The role of amplitude modulation can be easily exemplified using a real field with a pure amplitude modulation such as the one used in our sine-modulation Ansatz of Sec.\,\ref{subsec:shallowssp_SIA}. Using a fully contrasted modulation with a wavelength matching the system length,  i.e. $\psit(x)=\sqrt{2\Nt}\cos(2\pi x/L)$,  we find 
$\int_0^L \psit(x)dx=0$, yielding $\psiu=0$, $\Nun=0$ and $\psio=\psit$, $\No=\Nt$. Therefore, a pure amplitude modulation may yield a state with no uniform component.
 
For the effect of phase modulations, one can consider  yet  another Cauchy-Schwarz inequality, which reads $|\int_0^L \psit(x)dx|\leq \int_0^L |\psit(x)|dx$. As the terms of the inequality match $\psiu L$ either for the full field (left) or for its mere amplitude (right), one sees that
the presence of phase modulations in $\psit$ yields a further decrease of $\Nun$ compared to the pure amplitude modulated case. 
To exemplify the striking effect of phase modulations, let's consider an exemplary wave function with a pure phase modulation making a full rotation over the system's length. This writes $\psit(x)=\sqrt{\Nt}\exp(i 2\pi x/L)$. We  then find
$\int_0^L \psit(x)dx=0$ such that $\psiu=0$, $\Nun=0$ and $\psio=\psit$, $\No=\Nt$. Therefore, a pure phase modulation can also transfer a state completely into its modulating field.

\section{Full expression of the resonant high-energy scattering response}
\label{appendix:Dirac}

Section~\ref{subsec:highscatt} describes the SSS's response to a high-energy scattering probe in the linear and weakly interacting regime. If the elementary excitations dictating the response can be well approximated to plane waves, Eq.\,\eqref{eq:SqIA} gives the DSF values, $\Sqo$, and Eq.\,\eqref{eq:SqIAr} its resonant value. In the main text, details on the derivation of the prefactor were skipped to focus on the information related to the state's wave function. In this appendix, I provide an additional description to derive the full expression of the resonant response, Eq.\,\eqref{eq:SqIAr}. 

Here, I consider again a system well described by an effective 1D model and confined along $x$ in a uniform box of size $L$, see also Sec.\,\ref{subsec:decomp}. In this case, the plane-wave basis is discrete,  with the allowed momentum values being $\hbar k_j=h j/L$, $j\in Z$. I note $\epsilon_j = \hbar^2 k_j^2/2m$ the free-particle (kinetic) energy associated to the plane wave of index $j$, with $m$ being the particle's mass. In the regime of validity of Eqs.~\eqref{eq:SqIA}-\eqref{eq:SqIAr}, the response of the system to a scattering probe is dictated by elementary excitations. If the scattering probe is of large-enough momentum $\hbar k$, then the elementary excitation yielding a response is well approximated by the plane wave of index $j \approx k L /(2\pi)$~\footnote{I note that the resonant character of the response is also discretized and possible only if $k L /(2\pi) \in Z$.}. The energy of this elementary excitation reads $\hbar\omega_j \approx \epsilon_j + \Nt V(k_j)$ where $\Nt V(k_j)$ is a mean-field interaction shift. In this case, the regime of high-energy scattering where the impulse approximation is valid is defined by $\Nt V(k_j) \ll \hbar\omega_j, \epsilon_j$. In the case of dipolar gases, the interaction term, $V(k)$, can in general depend on $k$.  
For the dipolar system considered here, effective expression of $V(k)$ has been derived assuming an effective 1D model with transverse wave functions $\xi(y,z)$ of Gaussian shapes, see refs.~\cite{Blakie:2020,Deuretzbacher:2010,Sinha:2007}. Remarkably, for large-enough $k$, $V(k)$ was found to recover a momentum-independent character with $V(k)\sim V_{\infty}$.  
In this regime and assuming $\Nt V_{\infty} \ll \hbar\omega_j, \epsilon_j$, the Dirac factor appearing in Eq.\,\eqref{eq:SqIA} can then be simply written as:  
\begin{eqnarray}
\label{eq:Diracf}
\delta(\hbar\omega-\hbar\omega_j) &=& \delta\left(\hbar\omega-\Nt V_{\infty}-\frac{h^2 j^2}{2mL^2}\right)\\
\label{eq:Diracf3}
&=& \frac{2\pi mL}{h^2 |k_{\omega}|}\delta\left(j_\omega-j\right).
\end{eqnarray}
with $j_\omega =\frac{L}{h}\sqrt{2m(\hbar\omega-\Nt V_{\infty})}$ and  $k_{\omega}=2\pi j_\omega/L$. The summation of the factors \,\eqref{eq:Diracf3} over $j$ gives the density of excited states at energy $\hbar\omega$, $\sum_j \delta(\hbar\omega-\hbar\omega_j)=\frac{2\pi mL}{h^2 k_{\omega}}$. It matches the density of free-particle states in 1D taken at the mean-field-shifted energy $\hbar\omega-\Nt V_{\infty}$.  Combining this expression with Eq.~\eqref{eq:SqIA} yields Eq.\,\eqref{eq:SqIAr} for the DSF value on resonance with the state $j$ ($k=k_j$, $\omega=\omega_j$). In conclusion, the dependency of the resonant scattering response on the probed state at high momentum is contained in the factor $S$ of Eq.\,\eqref{eq:SqIAr} whereas the density-of-state factor yields an additional momentum dependence.

\section{Density contribution in the drop-array Ansatz}
\label{appendix:ndrop}
In Sec.\,\ref{subsec:DAA_dens}, an approximate expression of the density profile in the drop array is derived  under the assumption that the drops have a wave function that is steep enough over the range set by their half-distance. In the present Appendix, I detail the considerations that enable to simplify the density profile $n(x)$ as Eq.\,\eqref{eq:nxd_sumjl_split}. We see how these considerations yield the formulation of the steep-$|\chi|$ condition as Eq.\,\eqref{eq:steep_chi}.  

As stated in the main text, 
a primary step is to differentiate the terms of the sum appearing in the general definition of the density, Eq.\,\eqref{eq:nxdrop_sumjl}, as a function of the spacing, $s=l-j$, between the two drops identified by the couple of sum indexes $(j,l)$. This yields the expression  Eq.\,\eqref{eq:nxdrop_sumjl_s} where the  
$s$-overlap functions, $\eta_s$, defined in Eq.\eqref{eq:etadef}, come into play. 

To simplify Eq.\,\eqref{eq:nxdrop_sumjl_s} into Eq.\,\eqref{eq:nxd_sumjl_split}, let us first focus on the $s=0$ contribution. Its overall value, $\ndrt(x)$, is given by Eq.~\eqref{eq:ndrt}, matching the sum of the individual drop density profiles.
As $|\chi(x)|$ is monotonically decreasing for $x>0$, if one assumes \textbf{(A)} $|\chi(d/2)|^2\ll |\chi(0)|^2$, one can neglect the contributions of all other drops at one drop $j$'s center. In this way, one finds that $\ndrt(x)$ is structured as described in the main text. It has maxima at each drop center, $x \approx jd$, approximately equaling $\eta_0(0)$. Additionally, it has minima in between the drop, at $x\approx d(j+1/2)$, with an approximate value of $2\eta_0(d/2)$, from the contributions of the two neighboring drops, which is furthermore negligible compared to the maximum value. 

Differently, in the case $s\neq 0$, the maximum amplitude of $\eta_s(x)$ is found at an intermediate $x$ in $[0,sd]$. Using the symmetry of $|\chi(x)|$  and its monotonic character for $x>0$, we deduce two properties: (i), $|\eta_s(x)|\leq |\chi(0)\chi(sd/2)|$ for any $x$;  
(ii), if $|\chi(0)\chi^*(sd)|\ll |\chi(sd/2)|^2$, the maximum amplitude of $\eta_s$  is found at $x\approx sd/2$ and it approximately equals $\eta_0(sd/2)$. By assuming \textbf{(B)}, $|\chi(0)\chi^*(d)|\ll |\chi(d/2)|^2$, one deduces from the properties (i) and (ii) that 
\begin{enumerate}
    \item  $|\eta_s(x)|\ll \eta_1(x), \eta_0(x)$, for all $x \in [-d/2,d/2]$ and $s\geq 2$;
    \item $\eta_1(x) \approx \eta_0(x)$ for $x\approx (j+1/2)d$, which is where $\eta_1$ is approximately maximum, and $\eta_1(x) \ll \eta_0(x)$ for $x\approx jd$. 
\end{enumerate} The result (1), combined with the above consideration on $\ndrt$, implies that all terms with $s\geq 2$ can be neglected in Eq.\,\eqref{eq:nxdrop_sumjl_s}, yielding Eq.\,\eqref{eq:nxd_sumjl_split}. The result (2) implies the dominant character of  the $s$=0-contribution to the density, and the deep-supersolid character of the drop array, $C\approx 1$. Result (2) also determines the exact strength of the density minima of the SSS, and thus the values of the density contrast and of the superfluid fraction, Eqs.\,\eqref{eq:Cdeep},\,\eqref{eq:fsfetasim}. 

In conclusion, the steep-$|\chi|$ assumption needed to draw the conclusion given in the main text combines the conditions \textbf{(A)} and \textbf{(B)}. The condition \textbf{(B)} rewrites as $|\chi(d)|/|\chi(d/2)|\ll |\chi(d/2)|/|\chi(0)|$, matching the first inequality of Eq.\,\eqref{eq:steep_chi}, which defines the steep-$|\chi|$ condition. The last inequality appearing in Eq.\,\eqref{eq:steep_chi}, $|\chi(d/2)|/|\chi(0)|\ll1$, is not exactly needed but it is sufficient to imply \textbf{(A)} to be satisfied. We thus choose the condition Eq.\,\eqref{eq:steep_chi} as the relevant one to have both conditions \textbf{(A)} and \textbf{(B)} simultaneously satisfied.    

In the last paragraph of this appendix, we finally 
illustrate the general considerations above, using the Gaussian-drop model \eqref{eq:chig}. Using this model, the $\eta_s$-functions take simple forms, given by Eq.\,\eqref{eq:etadefg}. By performing a simple spatial derivation, \begin{eqnarray}
\frac{d\eta_s}{dx}(x)=-\frac{2x-sd}{\lambda^2 d^2}\cdot \eta_s(x),
\end{eqnarray}
and one finds that the maximum amplitudes of $\eta_s(x)$  are indeed found at $x=sd/2$ and equaling $\eta_0(sd/2)$.  
The $\ndrt(x)$ minima are similarly found at $x=d(j+1/2)$ and equaling $2\eta_0(d/2)$. Combining these results with Eq.\,\eqref{eq:etadefg} implies that the exponential factor $\exp(-1/4\lambda^2)$ fixes the ratio of the maximum amplitudes of $|\eta_{s}|$ and $|\eta_{s-1}|$. It also gives the ratio between the minima and maxima of $\ndrt(x)$ up to a factor 2. Finally, the steep-$|\chi|$ condition \eqref{eq:steep_chi} also writes as a function of this same factor, simply as $\exp(-1/4\lambda^2) \ll 1$. Therefore, under the steep-$|\chi|$ condition, in the Gaussian-drop Ansatz, all $s$>1 contributions are indeed found to be negligible and $\nov$ only compete with $\ndrt$ at the intermediate positions $x\approx d(j+1/2)$. This example corroborates the general considerations made above and given in the main text.

\end{document}